\documentclass[12pt]{article}
\usepackage{amssymb,slashed,latexsym,ifpdf,amsmath}
\def\rme{{\rm e}}

\def\K{K\"{a}hler}
\newsavebox{\uuunit}
\sbox{\uuunit}
    {\setlength{\unitlength}{0.825em}
     \begin{picture}(0.6,0.7)
        \thinlines
        \put(0,0){\line(1,0){0.5}}
        \put(0.15,0){\line(0,1){0.7}}
        \put(0.35,0){\line(0,1){0.8}}
       \multiput(0.3,0.8)(-0.04,-0.02){12}{\rule{0.5pt}{0.5pt}}
     \end {picture}}

\newcommand{\bbox}{\lower.2ex\hbox{$\Box$}}
\csname @addtoreset\endcsname{equation}{section}

\ifx\pdfoutput\undefined
   \pdffalse
   \usepackage{cite}
 \else
   \pdfoutput=1
   \pdftrue
  \usepackage[pdftex]{hyperref}
\fi

\newcommand{\rf}[1]{(\ref{#1})}

\def\aD3{{\overline {\rm D3}}}
\def\be{\begin{equation}}
\def\ee{\end{equation}}
\def\ba{\begin{array}}
\def\ea{\end{array}}
\def\bea{\begin{eqnarray}}
\def\eea{\end{eqnarray}}

\def\ib{{\bar \imath}}
\def\jb{{\bar \jmath}}
\def\K{K{\"a}hler}
\newcommand{\lp}{\left(}
\newcommand{\ls}{\left[}
\newcommand{\rp}{\right)}
\newcommand{\rs}{\right]}
\newcommand{\nn}{\nonumber}
\newcommand{\mL}{\mathcal{L}}

\begin{document}

\begin{titlepage}
\begin{flushright}
TUW-15-15
\end{flushright}
\vspace{.5cm}
\begin{center}
\baselineskip=16pt

{\Large {\bf De Sitter Supergravity Model Building}}

\

\

  {\large Renata Kallosh$^1$} and {\large   Timm Wrase$^2$} \vskip 0.8cm
{\small\sl\noindent
$^1$ SITP and Department of Physics, Stanford University, Stanford, CA
94305 USA \\\smallskip
$^2$Institute for Theoretical Physics, TU Wien, A-1040 Vienna, Austria}

\vskip 3cm
\begin{center}
{\bf Abstract}
\end{center}
\

\end{center}
{\small We present the explicit de Sitter supergravity action describing the interaction of supergravity with an arbitrary number of chiral and vector multiplets as well as one nilpotent chiral multiplet.  The action has a non-Gaussian dependence on the auxiliary field of the nilpotent multiplet, however, it can be integrated out for an arbitrary matter-coupled supergravity. The general supergravity action with a given \K\, potential $K$,  superpotential $W$ and vector matrix $f_{AB}$ interacting with a nilpotent chiral multiplet consists of the standard supergravity action defined by  $K$,  $W$ and  $f_{AB}$  where the scalar in the nilpotent multiplet has to be replaced by a bilinear combination of the fermion in the nilpotent multiplet divided by the Gaussian value of the auxiliary field. All additional contributions to the action start with terms quartic and higher order in the fermion of the nilpotent multiplet. These are given by a simple universal closed form expression.

 }


\

\vspace{2mm} \vfill \hrule width 3.cm
{\footnotesize \noindent e-mails: kallosh@stanford.edu, timm.wrase@tuwien.ac.at
 }
\end{titlepage}
\addtocounter{page}{1}
\newpage

\section{Introduction}

A supergravity action, including fermion interactions, with non-linearly realized spontaneously broken local supersymmetry was derived in \cite{Bergshoeff:2015tra, Hasegawa:2015bza}. In case of pure supergravity without scalars it was shown in \cite{Bergshoeff:2015tra} that de Sitter vacua form a simple set of classical solutions of this theory. For a long time only anti-de Sitter supergravity without scalars was known  \cite{Townsend:1977qa}. De Sitter vacua are also natural 
 for the theory with a single chiral multiplet, constructed in \cite{Hasegawa:2015bza}. The presence of a 
 Volkov-Akulov (VA) type non-linearly realized supersymmetry \cite{Volkov:1973ix} in supergravity models \cite{Bergshoeff:2015tra, Hasegawa:2015bza} leads to a natural uplifting of the vacuum energy in all these models. This is in full agreement with the string theory realization of the KKLT de Sitter vacua that use an anti-D3 brane \cite{Kachru:2003aw}. The construction in \cite{Kachru:2003aw} was  recently described as a supersymmetric realization of the KKLT uplifting with account of the fermions on the world-volume of the anti-D3 brane \cite{Bergshoeff:2015jxa}. Furthermore it was shown in \cite{Kallosh:2015nia} how this setup can be realized in explicit warped string compactifications.

The purpose of this paper is to derive general and explicit supergravity models with chiral and vector multiplets  interacting with a nilpotent chiral multiplet based on  the superconformal formulation of this theory given in  \cite{Ferrara:2014kva}. This strategy  was already used successfully in \cite{Bergshoeff:2015tra, Hasegawa:2015bza} where full actions with fermions  in models with a nilpotent multiplet and without matter  multiplets or with a single chiral multiplet were derived. Meanwhile in \cite{Tyutin} it was proposed how to derive supergravities with a nilpotent multiplet for general classes of models with any number of chiral and vector multiplets and with generic $K$,  $W$ and  $f_{AB}$. It was shown that  one can use the same method as in \cite{Bergshoeff:2015tra}, namely to perform a non-Gaussian  integration of the auxiliary field resulting in a closed form action. A complete action in the unitary gauge was presented in \cite{Tyutin}  for general matter coupling.  Here we will derive the complete action with fermions and with local supersymmetry for general matter coupling.  

The general action without a nilpotent multiplet is well known and we will use here the framework presented in \cite{Freedman:2012zz}. Originally the corresponding general supergravity-Yang-Mills  action was derived from the superconformal theory in \cite{Cremmer:1982wb,Kugo:1982mr,Kallosh:2000ve}.

The interest in a nilpotent chiral multiplet  satisfying the nilpotency condition in supergravity  in applications to cosmology was initiated in \cite{Antoniadis:2014oya} for the VA-Starobinsky model. For a general supergravity, interacting with the VA model, the superconformal construction was presented in \cite{Ferrara:2014kva}. It has been shown in \cite{Dudas:2015eha} that the approach of using the constrained curvature superfield is dual to the VA model coupled to supergravity.
It was also shown recently that one can use a complex linear goldstino superfield and build  de Sitter supergravity \cite{Kuzenko:2015yxa}.

This interest in these new supergravity models was increased  by the fact that they facilitate the construction of early universe inflationary models compatible with the data and the construction of de Sitter vacua for explaining dark energy and supersymmetry breaking, see for example \cite{Kallosh:2014via}  and references therein.
Such supergravity models in application to cosmology require mostly the knowledge of the bosonic action of the theory, where the rules are very simple: for the complete models  one has to construct  the standard supergravity action defined by  $K$,   $W$ and  $f_{AB}$. Once the action is known, one has to take only the bosonic part of it and, moreover, set the scalar field in the nilpotent multiplet to zero since the scalar in the nilpotent multiplet, representing the VA theory, is a fermion bilinear. If the scalar in the nilpotent multiplet is $z^1$, then the rule for obtaining the bosonic action is 
\bea\label{eq:Lb1}
e^{-1} { \cal L}_{\rm bosonic} = e^{-1} { \cal L}^{\rm book} [ K(z^\alpha, \bar z^{\bar \alpha}), W(z^\alpha) , f_{AB}(z^\alpha)]\Big |_{z^1=\chi^\alpha=\chi^{\bar \alpha} = \lambda^A =\psi_\mu =0}\, , \quad \alpha=1,...,n\,,
\eea
for a given choice of \K\, potential, superpotential and vector metric. Here  ${ \cal L}^{\rm book}$ is the standard supergravity action for a given choice of  $K$,   $W$ and  $f_{AB}$, see also next section for details. 

However, the knowledge of the bosonic action may not be sufficient for studies of reheating of the universe in these models, as well as for studies of particle physics, where the role of the fermions is important. A complete action including fermions,  with the bosonic part given in \rf{eq:Lb1}, will be derived here, following the proposal   in \cite{Tyutin}.

\section{Supergravity action with a nilpotent multiplet}\label{sec:result}
The general supergravity Lagrangian for an arbitrary number of chiral and vector multiplets coupled to supergravity is given for example in the book \cite{Freedman:2012zz} in equations (18.6)-(18.19) and takes the form 
\be\label{eq:Lb}
e^{-1} \mathcal{L}^{\rm book} = \mL_{\rm kin} -V + \mL_{\rm m} +\mL_{\rm mix} + \mL_{\rm 4f}\,,
\ee
where $\mL_{\rm kin}$ contains all the terms with spacetime derivatives of the fields. All terms on the right-hand-side of eq. \rf{eq:Lb} are defined in   \cite{Freedman:2012zz}. The Lagrangian is a function of all the physical fields $\mathcal{L}^{\rm book}=\mathcal{L}^{\rm book}(e_\mu^a,\psi_\mu,z^\alpha,\bar{z}^{\bar \alpha}, \chi^\alpha,\chi^{\bar \alpha},A_\mu^A,\lambda^A)$. 

We are interested in the case in which one of the chiral superfields in the supergravity action is nilpotent. Without loss of generality we choose the first one which leads to the constraint \cite{Bergshoeff:2015tra}: $z^1 = \frac{(\chi^1)^2}{2F^1}$  (and likewise $\bar{z}^{\bar 1} = \frac{(\chi^{\bar 1})^2}{2 \bar F^{\bar 1}}$).\footnote{We are using the conventions of \cite{Freedman:2012zz} but we will set $\kappa=1$ and we define the short-hand notations $(\chi^1)^2 = \bar{\chi}^1P_L \chi^1$ and $(\chi^{\bar 1})^2 = \bar{\chi}^{\bar 1} P_R\chi^{\bar 1}$.} What is the corresponding Lagrangian in this case? Clearly we cannot just plug $z^1 = \frac{(\chi^1)^2}{2F^1}$ into the on-shell Lagrangian \eqref{eq:Lb} since we do not know what $F^1$ is in the general supergravity model. What we actually need to do is to start with an off-shell supergravity Lagrangian and then integrate out all the $F^\alpha$ to get the correct on-shell Lagrangian.

The corresponding off-shell supergravity action for an arbitrary number of chiral and vector multiplets coupled to gravity is given by
\bea\label{eq:Loffshell}
e^{-1} { \cal L}_{\rm off-shell}= (F^\alpha - F^{\alpha } _G  ) g_{\alpha \bar \alpha }    (\bar F^{\bar \alpha} - \bar F^{\bar \alpha } _G   ) + e^{-1} { \cal L}^{\rm book}\,,
\eea
where
\be\label{eq:F}
F^{\alpha } _G = -\rme^{\frac{K}{2}} g^{\alpha \bar \beta} \overline\nabla _{\bar \beta }\overline{W} + (F_G^{\alpha})^f\,,
\ee
with
\bea\label{eq:Ff}
(F_G^{\alpha})^f &=& \frac12 \Gamma^\alpha_{\beta \gamma} \bar \chi^\beta \chi^\gamma +\frac14 \bar f_{AB \bar{\beta}} g^{\bar \beta \alpha} \bar{\lambda}^A P_R \lambda^B\,.
\eea
Alternatively, we could have started with the off-shell supergravity action in \cite{Kugo:1982mr}, eqs. (25)-(33),  which is given in the form with all auxiliary fields not integrated out. This set up was used in \cite{Hasegawa:2015bza}. When the auxiliary fields $A_\mu$ and $F^0$ are integrated out, the remaining action still has all auxiliary fields $F^{\alpha}$  as shown in our eq. \rf{eq:Loffshell}.

Now since the Lagrangian ${\cal L}^{\rm book}$ does depend on $z^1 = \frac{(\chi^1)^2}{2F^1}$ it will also depend on $F^1$ and therefore it seems prohibitively difficult to explicitly integrate it out. However, using a proposal made in \cite{Tyutin} which includes a simplifying assumption about the form of {\K} potential we will derive the explicit on-shell Lagrangian in the next section. In particular, we assume that the K\"ahler potential depends only on the product $z^1 \bar{z}^{\bar 1}$ so that we have \footnote{Since $(z^1)^2=(\bar{z}^{\bar 1})^2=0$ the only other term that could arise in the fully general {\K} potential is $z^1 f(z^i,\bar z^\ib) +\bar z^{\bar 1} \bar f(z^i,\bar z^\ib)$. Note that linear terms  in $z^1$ ($\bar z^{\bar 1}$) that are multiplied by a holomorphic (anti-holomorphic) function can be removed by a {\K} transformation.}
\be\label{eq:K}
K(z^i,\bar{z}^{\ib}, z^1\bar z^{\bar 1}) = K_0(z^i,\bar{z}^{\ib}) + z^1 \bar{z}^{\bar 1} g_{1 \bar 1}(z^i,\bar z^\ib)\,.
\ee
We also expand the holomorphic superpotential $W$ and the holomorphic gauge kinetic function $f_{AB}$ as follows
\bea
W(z^1,z^i) &=& W_0(z^i) + z^1 \,W_1 (z^i) \equiv g(z^i) + z^1 f(z^i) \,,\cr
\cr
f_{AB}(z^1,z^i) &=& f_{AB0}(z^i) +z^1 \, f_{AB1}(z^i)\,.
\eea 
With this assumption we will integrate out $F^1$ in the next section \ref{sec:details}. This involves taking into account a non-Gaussian dependence of the action on the auxiliary field $F^1$. This gives us the generic supergravity Lagrangian for a nilpotent chiral superfield coupled to an arbitrary number of chiral and vector multiplets:
\bea\label{eq:L10}
\boxed{e^{-1} { \cal L}_{\rm final} = \Big [ e^{-1} { \cal L}^{\rm book} \Big ]_{z^1=\frac{(\chi^1)^2}{2 f^1}} -   \frac{(\chi^1)^2\, \,  (\chi^{\bar 1})^2}{ 4 g_{1 \bar 1} (f^1\bar f^{\bar 1})^2} \left|  g_{1\bar 1} \frac{\Box(\chi^1)^2}{2 f^1}+ B^{1}  \right |^2 }\,\,\,
\eea
with
\bea\label{eq:B1}
f^1 &\equiv &  \frac{1}{g_{1 \bar 1}} \lp -\rme^{\frac{K_0}{2}}\,\overline{W}_{\bar 1} +\frac{1}{4 } \bar f_{AB \bar{1}} \bar{\lambda}^A P_R \lambda^B \rp\,,\nn\\[2mm]
B^{1} & \equiv & e^{-1} { \delta { \cal L}^{\rm book}\over \delta \bar z^{\bar 1}} |_{z^1= \bar z^{\bar 1}=0}\, .
\eea
Note that the last term in equation \rf{eq:L10} has already the maximal power of the (undifferentiated) spinor $\chi^1$ so that we can drop all terms in $B^1$ that contain $\chi^1$ or $\chi^{\bar 1}$. We explicitly give the relevant part of $B^1$ for two examples in section \ref{sec:examples}.

One can see that the complete Lagrangian \eqref{eq:L10} has the original form with $z^1$ replaced by its `Gaussian value' and there are some additional terms that are at least quadratic in $\chi^1 \chi^{\bar 1}$. On the basis of this expression we may conclude that the action with $K$, $W$ and $f_{AB}$ generic (but with $K$ dependence on $z^1 \bar z^{\bar 1}$) is given by the standard action with the following modifications:
 
After the standard action for $K$, $W$ and $f_{AB}$ is presented as a function of $ z^1$ and $\bar z^{\bar 1}$
\begin{enumerate}
\item Replace $ z^1$ by $z^1 =\frac{(\chi^1)^2}{2 f^1}$ and likewise for the conjugate $\bar z^{\bar 1}= \frac{(\chi^{\bar 1})^2}{2 \bar f^{\bar 1}}$.
\item Add the quartic and higher order in spinors terms shown in the second entry in eq. \rf{eq:L10}, where $B^{1}$ is given in \eqref{eq:B1} and more explicitly for two examples in section \ref{sec:examples}.
\end{enumerate}

\section{Derivation of the  matter-coupled supergravity action with a nilpotent multiplet}\label{sec:details}
We start with the off-shell action \rf{eq:Loffshell} and we would like to integrate out the auxiliary fields $F^\alpha$. The dependence on 
 $F^1$ is non-Gaussian since the Lagrangian depends on $z^1= {(\chi^1)^2\over 2 F^1}$. However, we can still trivially integrate out all the other $F^i$, $i=2,\ldots,n$ and get
\be
g_{i \bar 1}(\bar F^{\bar 1} -\bar F^{\bar 1}_G ) + g_{i \jb}(\bar F^{\jb} -\bar F^{\jb}_G ) =0\,.
\ee
The sub-matrix $g_{i\jb}$ is invertible in order to have a non-degenerate kinetic term for the $z^i$. Thus we find
\be
(\bar F^{\jb} -\bar F^{\jb}_G ) = -(g_{\jb i})^{-1}  g_{i \bar 1}(\bar F^{\bar 1} -\bar F^{\bar 1}_G )\,,
\ee
and the Lagrangian after integrating out the $F^i$ becomes
\be
e^{-1} { \cal L}_{\rm off-shell}= (F^1 - F^{1 } _G  ) (g_{1 \bar 1 } -g_{1\jb} (g_{\jb i})^{-1}  g_{i \bar 1})   (\bar F^{\bar 1} - \bar F^{\bar 1 } _G   ) + e^{-1} { \cal L}^{\rm book}\,.
\label{action} \ee
Here $e^{-1} { \cal L}^{\rm book}$ is defined as an expansion in $z^1, \bar z^{\bar 1}$ 
\be\label{eq:Lbook}
e^{-1} { \cal L}^{\rm book} = \bar z^{\bar 1} A^1 z^1 +B^1\bar z^{\bar 1}  + \bar B^1 z^1 + C^1 \,.
\ee
In order to solve for $F^1$ we follow the approach described in \cite{Tyutin}. Using the {\K} potential in equation \rf{eq:K} which immediately implies that
\be\label{eq:g}
g_{i \bar1} = z^1 \partial_i g_{1\bar 1}\,, \qquad g_{1 \ib} = \bar z^{\bar 1} \partial_{\ib} g_{1 \bar 1}\,,
\ee
we rewrite the first term in the Lagrangian \rf{action} as follows
\bea\label{eq:L}
 (F^1 - F^{1 } _G  ) g_{1 \bar 1}(\bar F^{\bar 1} - \bar F^{\bar 1 } _G   ) - (F^1 - F^{1 } _G  )z^1 \bar{z}^{\bar 1} (\partial_{\jb} g_{1 \bar 1})  (K_{0,\jb i})^{-1} (\partial_i g_{1 \bar 1})   (\bar F^{\bar 1} - \bar F^{\bar 1 } _G   )\,,
\eea
where we used that $(g_{\jb i})^{-1}=  (K_{0,\jb i})^{-1}-z^1 \bar z^{\bar 1}  (K_{0,\jb l})^{-1}  (K_{0,i \bar m})^{-1} (\partial_l \partial_{\bar m} g_{1 \bar 1})$ and the fact that $(z^1)^2=0$. 

When solving for $F^1$, following \cite{Bergshoeff:2015tra}, one finds that $F^1 = F^1_{G}$ up to powers of $\chi^1$ (see eqn. (A.24) in \cite{Bergshoeff:2015tra}). Since the second term in equation \rf{eq:L} has already the maximal power of $\chi^1$ because $z^1 \bar{z}^{\bar 1} \propto (\chi^1)^2 (\chi^{\bar 1})^2$ we find that the second term does not contribute to the action at all. We show this explicitly in appendix \ref{app:deltas}. Thus, we find that we are left with 
\be
e^{-1} { \cal L}_{\rm off-shell}= (F^1 - F^{1 } _G  ) g_{1 \bar 1 }   (\bar F^{\bar 1} - \bar F^{\bar 1 } _G   ) + e^{-1} { \cal L}^{\rm book}\,.
\label{action1} \ee
Now following \cite{Tyutin} we want to separate all the explicit $z^1$, $\bar z^{\bar 1}$ dependence in the first  term of \rf{action1} from the dependence on $F^1$ and $\bar F^{\bar 1}$. This will then bring the Lagrangian into a form in which we can easily integrate out $F^1$ following the procedure developed in \cite{Bergshoeff:2015tra}. To do that we define the following expansion coefficients that are independent of $z^1$ and $\bar z^{\bar 1}$
\be\label{eq:F1exp}
F^1_G = F^1_{G0} + F^1_{G1} z^1 + F^1_{G\bar 1} \bar z^{\bar 1} + F^1_{G1\bar 1} z^1 \bar z^{\bar 1}\,,
\ee
and likewise for $\bar F^{\bar 1}_G$. We present some details on the moduli space geometry for our models in appendix \ref{app:C}. Using it, a straightforward calculation, given in appendix \ref{app:F1}, shows that $F^1_{G \bar 1} = \bar F^{\bar 1}_{G 1}=0$ for the {\K} potential given in equation \rf{eq:K}.

Now let us look at the term $(F^1 - F^{1 } _G  )g_{1 \bar 1} (\bar F^{\bar 1} - \bar F^{\bar 1 } _G  )$. We can absorb the $z^1$ and $z^1 \bar z^{\bar 1}$ dependent terms of $F^{1}_G$ into $F^1$ by defining 
\be
F^{'1} = F^1 - F^1_{G1} z^1 - z^1\bar z^{\bar 1} F^1_{G1\bar 1}\,.
\ee
This does not affect the nilpotency condition that fixes $z^1=\frac{(\chi^1)^2}{2F^1}$ since
\be
z^1 = \frac{(\chi^1)^2}{2 F^1} = \frac{(\chi^1)^2}{2 F^{'1} } = \frac{(\chi^1)^2}{2 F^1} \lp 1 + \frac{z^1}{2 F^1} \lp F^1_{G1} + \bar z^{\bar 1} F^1_{G1\bar 1}\rp \rp\,.
\ee
Here we used that $(\chi^1)^2 z^1 \propto (\chi^1)^2 (\chi^1)^2 = 0$. Likewise we can define 
\be
\bar F^{' \bar 1} = \bar F^{\bar 1} - \bar F^{\bar 1}_{G\bar 1}\bar z^{\bar 1} - z^1\bar z^{\bar 1} \bar F^{\bar 1}_{G1\bar 1}\,.
\ee 
This leads to 
\be
(F^1 - F^{1 } _G  )g_{1 \bar 1} (\bar F^{\bar 1} - \bar F^{\bar 1 } _G  ) = (F^{'1} - F^{1}_{G0}) g_{1 \bar 1}(\bar F^{' \bar 1} - \bar F^{\bar 1 } _{G0})\,.
\ee
Putting everything together we can now rewrite the Lagrangian \eqref{action1} as
\bea\label{eq:L2}
e^{-1} { \cal L} &=& (F^{'1} - F^{1 } _{G0}) g_{1 \bar 1} (\bar F^{'\bar 1} - \bar F^{\bar 1 } _{G0}   ) + \bar z^{\bar 1} A z^1  + B \bar z^{\bar 1}+\bar { B} z^1+  C^1\,.\qquad
\eea
We now introduce $F=F^{'1} \sqrt{g_{1\bar 1}}$, $F_{G0} = F^1_{G0}  \sqrt{g_{1\bar 1}}$, $z=z^1/ \sqrt{g_{1\bar 1}}=\frac{(\chi^1)^2}{2F}$ and similarly for the conjugated quantities. We also define $A=g_{1 \bar 1} A^1 $, $B=\sqrt{g_{1 \bar 1}} B^1 $, $C= C^1$ to bring the Lagrangian into the form \footnote{Note that our definition of $C$ is different from the one used in \cite{Bergshoeff:2015tra}.}
\bea\label{eq:L3}
e^{-1} { \cal L}_{\rm off-shell} &=& (F- F_{G0})(\bar F - \bar F_{G0}   ) + \bar z A z+B \bar z + \bar B z + C\,.
\eea
Now we can solve the equation for $F$ that is given by
\be
\frac{\delta \mathcal{L}(z,\bar z,F,\bar F)}{\delta F} + \frac{\partial z }{\partial F} \frac{\delta \mathcal{L}(z,\bar z,F,\bar F)}{\delta z} =\frac{\delta \mathcal{L}(z,\bar z,F,\bar F)}{\delta F}-\frac{ z }{ F} \frac{\delta \mathcal{L}(z,\bar z,F,\bar F)}{\delta z} =0\,.
\ee
This was done for the Lagrangian \rf{eq:L3} in the paper \cite{Bergshoeff:2015tra} (see in particular appendix A.5). The resulting on-shell Lagrangian is given by
\bea\label{eq:L4}
e^{-1} { \cal L}_{\rm on-shell} &=& \Big [   \bar z\, A\,  z+ z \bar B + B\bar z + C   -  \frac{1}{ F_{G0}\bar F_{G0}}\lp \bar z A z+\bar z B \rp \lp z \bar A \bar z+z \bar B \rp \Big ]_{z=\frac{(\chi^1)^2}{2 F_{G0}}} \,.\qquad
\eea
Expressing this in our original variables we have
\bea\label{eq:L5}
e^{-1} { \cal L}_{\rm on-shell} &=& \Big [   \bar z^{\bar1}\, A^1 \,  z^1+ z^1 \bar B^1  + B^1 \bar z^{\bar 1}+ C^1   -  \frac{1}{g_{1 \bar 1} F^1_{G0}\bar F^{\bar 1}_{G0}} \left|\bar z^{\bar 1} A^1 z^1+\bar z^{\bar 1} B^1\right |^2 \Big ]_{z^1=\frac{(\chi^1)^2}{2 F^1_{G0}}}\cr
 &=& \Big [ e^{-1} { \cal L}^{\rm book}  -  \frac{1}{g_{1 \bar 1} F^1_{G0}\bar F^{\bar1}_{G0}} \left|\bar z^{\bar 1} A^1 z^1+\bar z^{\bar 1} B^1\right |^2 \Big ]_{z^1=\frac{(\chi^1)^2}{2 F^1_{G0}}} \,.\qquad
\eea
To make it fully explicit that the `new' terms in the Lagrangian contain the maximal power $(\chi^1)^2 (\chi^{\bar 1})^2$ of the undifferentiated spinor $\chi^1$, we can rewrite the Lagrangian as
\bea\label{eq:L8}
e^{-1} { \cal L}_{\rm final} = \Big [ e^{-1} { \cal L}^{\rm book} \Big ]_{z^1=\frac{(\chi^1)^2}{2 F^{1}_{G0}}} - \frac{(\chi^1)^2\, \,  (\chi^{\bar 1})^2 }{4 g_{1 \bar 1} (F^1_{G0}\bar F^{\bar1}_{G0})^2} \left| A^1 \frac{(\chi^1)^2}{2 F^1_{G0}}+ B^1  \right |^2 \,,
\eea
where the explicit expression for $F^1_{G0}$ is derived in appendix \ref{app:F1} and given by
\be
F^1_{G0} =  \frac{1}{g_{1 \bar 1}} \lp- \rme^{\frac{K_0}{2}}\,\overline{W}_{\bar 1}+\frac{1}{2} (\partial_i g_{1 \bar 1})\bar \chi^{1} \chi^i +\frac{1}{4 } \bar f_{AB \bar{1}} \bar{\lambda}^A P_R \lambda^B \rp\,.
\ee
Note that $A^1$ in the above expression has to act with two derivatives on $(\chi^1)^2$ in order to give a non-zero expression. In the Lagrangian $\mL^{\rm book}$ in \eqref{eq:Lbook} the relevant part of the $\bar z^{\bar 1} A^1 z^1$ term which appears in the second term in \rf{eq:L8} is
\be
\bar z^{\bar 1}  g_{1\bar 1}g^{\mu\nu}  \partial _\mu  \partial _\nu z^1\equiv  \bar z^{\bar 1}  g_{1\bar 1}\Box z^1\,.
\ee
 Likewise, from $B^1$ and $F^1_{G0}$ only parts contribute that are independent of the undifferentiated $\chi^1$ and $\chi^{\bar 1}$ since they appear in the above action with the highest power of the spinor $\chi^1$. We denote these parts by $b^1$ and $f^1$, respectively. We explicitly spell out $b^1$ for two examples in section \ref{sec:examples} and $f^1$ in appendix \ref{app:F1}. Taking all this into account the Lagrangian \rf{eq:L8} simplifies to our final result given in equation \eqref{eq:L10} above.

\section{Examples}\label{sec:examples}
In this section we discuss two examples of our general result. First we discuss the simplest case of pure supergravity without chiral and vector multiplets. The action in this case was first derived in \cite{Bergshoeff:2015tra} and \cite{Hasegawa:2015bza}. Then we discuss a pretty general case in which we allow for an arbitrary number of chiral and vector multiplets and only make the assumption that the gauge kinetic function $f_{AB}$ and the moment maps $\mathcal{P}_A$ are independent of $z^1$.

\subsection{Pure dS supergravity}
For the case of a single nilpotent chiral superfield coupled to supergravity the most general {\K} and superpotential are given by
\bea\label{eq:KWsimple}
K = z^1 \bar z^{\bar 1}\, , \quad W =  g+f z^1\,,
\eea
with $g$ and $f$ complex constants. The corresponding action was derived in \cite{Bergshoeff:2015tra} and \cite{Hasegawa:2015bza} in two different superconformal gauges. The one we are using here is based on the framework in  \cite{Freedman:2012zz}.

Here we show how the action follows from our general answer given in \eqref{eq:L10}. For the simple {\K} and superpotential given in \eqref{eq:KWsimple} we find that $F^1_{G0}=-\bar{f}$. This leads to the action
\bea\label{eq:puredS}
e^{-1} { \cal L}_{\rm pure\, dS} = \Big [ e^{-1} { \cal L}^{\rm book} \Big ]_{z^1=-\frac{(\chi^1)^2}{2\bar f}} - \frac{ (\chi^1)^2\, \,  (\chi^{\bar 1})^2}{4 |f|^4} \left|  -\frac{\Box(\chi^1)^2}{2 \bar f}+ b^1  \right |^2 \,,
\eea
where the explicit expression for $b^1$ can be read off from the general Lagrangian in eqns. (18.6)-(18.19) in \cite{Freedman:2012zz}. In particular, $\alpha, \beta$ only take the value 1 and we want to extract all terms that are linear in $\bar{z}^{\bar 1}$ and independent of $z^1$ as described in equation \eqref{eq:B1}. Furthermore, we can drop all terms that contain the undifferentiated spinor $\chi^1$ or $\chi^{\bar 1}$ (after potentially integrating by parts). This leaves us with only three terms, one coming from the scalar potential $V$, one  coming from the mass term of the gravitino and lastly one term which comes from the term $+\frac{1}{\sqrt{2}} \bar \psi_\mu \slashed{\partial}\bar z^{\bar 1} \gamma^\mu \chi^1$ in the action after partial integration. The resulting answer is
\be\label{eq:B1puredS}
b^1 = 2 g \bar f +\frac12 \bar f \bar \psi_\mu P_L  \gamma^{\mu\nu} \psi_\nu- \frac{1}{\sqrt{2}} \bar \psi_\mu \gamma^\nu \gamma^\mu \partial_\nu \chi^1 \,. 
\ee

\subsection{De Sitter  supergravity coupled to chiral and vector superfields}
In this subsection we spell out the action in the case of an arbitrary number of chiral and vector multiplets but we make the simplifying assumption that the gauge kinetic function $f_{AB}$ and the moment maps $\mathcal{P}_A$ are independent of $z^1$ (and $\bar z^{\bar 1}$). The action is given by
\bea\label{eq:L11}
e^{-1} { \cal L}_{\rm final} = \Big [ e^{-1} { \cal L}^{\rm book} \Big ]_{z^1=\frac{(\chi^1)^2}{2 f^{1}}} -  \frac{(\chi^1)^2\, \,  (\chi^{\bar 1})^2}{4 g_{1 \bar 1} (f^1\bar f^{\bar1})^2} \left|  g_{1\bar 1} \frac{\Box(\chi^1)^2}{2 f^{1}}+ b^1  \right |^2 \,,
\eea
with
\bea
f^1 &=&  -\frac{ \rme^{\frac{K_0}{2}}\,\overline{W}_{\bar 1}}{g_{1 \bar 1}} \,,\nn\\[2mm]
b^1 &=& \rme^{K_0}\ls 2 W_0 \overline{W}_{\bar 1} +(K_{0,i\jb})^{-1} \lp D_{0,i}W_0\rp\lp \frac{\partial_\jb g_{1 \bar 1}}{g_{1 \bar 1}}\overline{W}_{\bar 1} -\bar{D}_{0,\jb} \overline{W}_{\bar 1}\rp\rs\cr
&&- \frac12 (\bar \chi^{\jb} \slashed{\partial}\chi^1) \ls 3 \partial_\jb g_{1 \bar 1}  +(\bar{z}^{\bar k}  \partial_{\bar k} -z^k \partial_k)  \partial_\jb g_{1\bar1} \rs\cr
&&+\frac12 \rme^{\frac{K_0}{2}} \overline{W}_{\bar 1} \bar \psi_\mu P_L  \gamma^{\mu\nu} \psi_\nu - \frac{1}{\sqrt{2}} \bar \psi_\mu \gamma^\nu \gamma^\mu \partial_\nu \chi^1 \big( g_{1 \bar 1} +(\partial_\jb g_{1\bar 1}) \bar z^\jb\big)\cr
&&+\frac14 \rme^{\frac{K_0}{2}} f_{ABj} \ \bar \lambda^A P_L \lambda^B \ (K_{0,j\bar k})^{-1} \lp(\partial_{\bar k} +K_{0,\bar k})  -\frac{\partial_{\bar k} g_{1 \bar 1}}{g_{1 \bar 1}} \rp \overline{W}_{\bar 1} \cr
&&-\frac12\rme^{\frac{K_0}{2}}  \ \bar{\chi}^\ib \chi^\jb \ \lp \bar{D}_{0,\ib} \bar{D}_{0,\jb} -\frac{\partial_\ib \partial_\jb g_{1 \bar 1}}{g_{1\bar1}}+(K_{0,k\bar l})^{-1} (K_{0,k\ib\jb}) \lp \frac{(\partial_{\bar l} g_{1 \bar 1})}{g_{1\bar1}}-\bar{D}_{0,\bar l}\rp \rp \overline{W}_{\bar 1} \cr
&&+\frac{1}{\sqrt{2}} \rme^{\frac{K_0}{2}} \bar{D}_{0,\ib} \overline{W}_{\bar 1} \ \bar \chi^\ib\, \gamma\cdot\psi \, ,
\eea 
where we defined $D_{0,i} = \partial_i + K_{0,i}$  and likewise $\bar{D}_{0,\ib} =  \partial_\ib + K_{0,\ib}$. This explicit expression for $b^1$ can again be read off from the action given in the book \cite{Freedman:2012zz}. In particular, the first line arises from the scalar potential $V$, the second line comes from the kinetic terms for the fermions and the third line contains the generalization of the terms that we already found above in equation \eqref{eq:B1puredS}. The fourth line comes from the gaugino mass term and the fifth from the mass term for the $\chi^i$. The last line contains a term coming from $\mathcal{L}_{\rm mix}$ in equation (18.18) in \cite{Freedman:2012zz}.

Note that this fairly general case includes the dS supergravity action coupled to a single chiral multiplet that was derived in \cite{Hasegawa:2015bza}.

\section{Summary}
We constructed the locally supersymmetric supergravity action for general  models with chiral and vector multiplets, presented  in eq. \rf{eq:L10}. It depends on all chiral multiplets, $z^\alpha$:   the nilpotent one $z^1$ and all other $z^i$ with $i=2,...,n$. Our  models are defined 
 by $K(z^\alpha, \bar z^{\bar \alpha})$,   $W(z^\alpha)$ and  $f_{AB}(z^\alpha)$ given by the following expressions 
\bea\label{eq:K1}
K(z^\alpha ,\bar{z}^{\bar \alpha }) &= & K_0(z^i,\bar{z}^{\ib}) + z^1 \bar{z}^{\bar 1} g_{1 \bar 1}(z^i,\bar z^\ib)\,,\cr
\cr
W(z^\alpha) &= & W_0(z^i) + z^1 W_1(z^i) \,,\cr
\cr
f_{AB}(z^\alpha) &=& f_{AB0}(z^i) +z^1 \, f_{AB1}(z^i)\,.
\eea 
Here the choice of $W$ and $f_{AB}$ is most general, since $(z^1)^2=0$, whereas the \K\, potential is assumed to depend on $z^1 \bar{z}^{\bar 1}$.   

Interesting features of de Sitter supergravity coupled to generic chiral and vector multiplets  and a nilpotent chiral multiplet  became clear after the complete action in eq. \rf{eq:L10} was derived:
\begin{enumerate}
  \item The bosonic action is the standard supergravity  action defined by $K(z^\alpha, \bar z^{\bar \alpha})$,   $W(z^\alpha)$ and  $f_{AB}(z^\alpha)$ which depend on all chiral multiplets. In this bosonic action one has to take $z^1=0$.
  
  \item The complete fermionic action up to terms quadratic in $\chi^1 \chi^{\bar 1}$ is given by the standard supergravity action defined by $K(z^\alpha, \bar z^{\bar \alpha})$,   $W(z^\alpha)$ and  $f_{AB}(z^\alpha)$, in which $z^1$ has to be replaced by  
\be
z^1 =  g_{1\bar 1}(z^i,\bar{z}^\ib) { (\chi^1)^2  \over 2 }   \lp -\rme^{\frac{K_0}{2}}\,\overline{W}_{\bar 1} (\bar z^\ib)  +\frac{1}{4 } \bar f_{AB \bar{1}}(\bar z^\ib) \bar{\lambda}^A P_R \lambda^B\rp^{-1} .
\label{repl}\ee
  
    \item The complete fermionic action to all orders in fermions is given by the standard supergravity action defined by $K(z^\alpha, \bar z^{\bar \alpha})$,   $W(z^\alpha)$ and  $f_{AB}(z^\alpha)$, in which $z^1$ has to be replaced as shown in eq. \rf{repl}. In addition to this, a term  quartic and higher order in fermions  has to be added to the action. It is given in closed form by the second term in eq. \rf{eq:L10}.
    
\item In the unitary gauge 
\be
\chi^1=0\,,
\ee
 the action reduces to  the standard supergravity action defined by $K(z^\alpha, \bar z^{\bar \alpha})$,   $W(z^\alpha)$ and  $f_{AB}(z^\alpha)$ taken at $\chi^1=z^1=0$ but with $F^1 \neq 0$:
 \be\label{eq:Funitary}
F^{1 }|_{z^1=\chi^1=0}  =\frac{1}{g_{1 \bar 1}(z^i,\bar{z}^\ib)}\lp -\rme^{\frac{K_0}{2}}  \overline{W}_{\bar 1}(\bar z^{\ib})  +\frac14 \bar f_{AB \bar{1}}(\bar z^\ib)  \bar{\lambda}^A P_R \lambda^B\rp \,.
\ee
 The non-linearly realized local supersymmetry of the action in the unitary gauge $\chi^1=0$ is broken. The extra terms due to $F^1$ from the  nilpotent multiplet  include the non-vanishing positive term in the potential 
\be
V=   e^{K_0(z^i,\bar{z}^{\ib})} |W_1(z^i)|^2 g^{1\bar 1}(z^i,\bar{z}^{\ib})  >0 \ ,
\label{Uplift}\ee
as well as some  fermionic terms in case that  $f_{AB }$ depends\footnote{In string theory constructions of the nilpotent multiplet via the anti-D3-brane one may argue that $f_{AB }$ does not depend on $z^1$  \cite{Kallosh:2015nia}.   In such case, the only effect of the nilpotent multiplet in the supergravity unitary gauge $\chi^1=0$ is the vacuum energy uplift \rf{Uplift}.} on $z^1$.
This is the basic feature of all de Sitter supergravity models, the leftover of the positive energy term even in the unitary local supersymmetry gauge in which the VA fermion is absent.   
\end{enumerate}

In conclusion, in this paper we have constructed a locally supersymmetric supergravity action for the case with a nilpotent multiplet and generic chiral and vectors multiplets. This creates a consistent framework for the investigation of phenomenological consequences of theories with nilpotent fields for particle physics and cosmology.

\section*{Acknowledgments}

We are grateful  to  E. Bergshoeff, S. Ferrara, D. Freedman,  A. Linde and   F. Quevedo
for stimulating  discussions.  The work of RK and TW is supported by the SITP, by the NSF Grant PHY-1316699 and  by the Templeton foundation grant `Quantum Gravity Frontiers'. TW thanks the Department of Physics of Stanford University and the Templeton foundation for the hospitality during a visit in which this work was initiated.

\appendix

\section{Vanishing of the extra term}\label{app:deltas}
In this section we show explicitly that the second term in \rf{eq:L} does not contribute to the final on-shell action.
We first recall that $F^1 z^1 = F^1 \frac{(\chi^1)^2}{2 F^1} = \frac{(\chi^1)^2}{2}$ is $F^1$ independent (and likewise for $\bar F^{\bar 1} \bar z^{\bar 1}$). This allows us to rewrite the second term in the Lagrangian \eqref{eq:L}, which due to the factor $z^1 \bar z^{\bar 1}$ can only depend on $F^1_{G0}$ and $\bar F^{\bar 1}_{G0}$ (see equation \eqref{eq:F1exp} for the definition)
\bea
&&-(F^1 - F^{1 } _{G0}) z^1 \bar{z}^{\bar 1} (\partial_{\jb} g_{1 \bar 1})  (K_{0,\jb i})^{-1} (\partial_i g_{1 \bar 1}) (\bar F^{\bar 1} - \bar F^{\bar 1 } _{G0}   )  \\
&&=  (\partial_{\jb} g_{1 \bar 1})  (K_{0,\jb i})^{-1} (\partial_i g_{1 \bar 1}) \lp -z^1 \bar{z}^{\bar 1} F^{1 } _{G0} \bar F^{\bar 1 } _{G0}  + z^1 \frac{(\chi^{\bar1})^2}{2} F^{1 }_{G0} + \bar z^{\bar 1} \frac{(\chi^1)^2}{2} \bar F^{\bar 1 }_{G0} - \frac{(\chi^1)^2 (\chi^{\bar 1})^2}{4} \rp\,. \nn
\eea
So we see that the explicit $F^1$, $\bar{F}^{\bar 1}$ dependence completely disappears. The above terms then become a correction to $\eqref{eq:Lbook}$ that of course also does not have an explicit dependence on $F^1$ and $\bar{F}^{\bar 1}$. In particular we find that the coefficient $A^1$, $B^1$, $\bar{B}^1$ and $C^1$ get the following extra contributions
\bea\label{eq:DABC}
\Delta A^1 &=& - \mathcal{M} F^{1 } _{G0} \bar F^{\bar 1 } _{G0}\,,\cr
\Delta B^1 &=& \mathcal{M} \frac{(\chi^1)^2}{2} \bar F^{\bar 1 }_{G0} \,, \cr
\Delta \bar B^1 &=& \mathcal{M} \frac{(\chi^{\bar 1})^2}{2} F^{1 }_{G0}\,, \cr
\Delta C^1 &=& -\mathcal{M} \frac{(\chi^1)^2 (\chi^{\bar 1})^2}{4} \,, 
\eea
with
\be
\mathcal{M} =(\partial_{\jb} g_{1 \bar 1})  (K_{0,\jb i})^{-1} (\partial_i g_{1 \bar 1})\,.
\ee
Then we can proceed as in section 3 and integrate out $F^1$. The final action expressed in our original variables takes the form (cf. eqn. \eqref{eq:L5})
\bea\label{eq:Lapp}
e^{-1} { \cal L}_{\rm on-shell} &=& \Big [   \bar z^{\bar1}\, (A^1+\Delta A^1) \,  z^1+ z^1 (\bar B^1 +\Delta \bar{B}^1) + (B^1 +\Delta B^1) \bar z^{\bar 1}+ (C^1+\Delta C^1)   \cr
&&-  \frac{1}{g_{1 \bar 1} F^1_{G0}\bar F^{\bar 1}_{G0}} \left|\bar z^{\bar 1} (A^1+\Delta A^1) z^1+\bar z^{\bar 1} (B^1+\Delta B^1)\right |^2 \Big ]_{z^1=\frac{(\chi^1)^2}{2 F^1_{G0}}}\cr
&=& \Big [ e^{-1} { \cal L}^{\rm book} +\bar z^{\bar1}\, \Delta A^1 \,  z^1+ z^1 \Delta \bar{B}^1 + \Delta B^1 \bar z^{\bar 1}+ \Delta C^1\cr
&&-  \frac{1}{g_{1 \bar 1} F^1_{G0}\bar F^{\bar 1}_{G0}} \left|\bar z^{\bar 1} A^1 z^1+B^1\bar z^{\bar 1}+ \bar z^{\bar 1} \Delta A^1 z^1+ \Delta B^1\bar z^{\bar 1} \right |^2 \Big ]_{z^1=\frac{(\chi^1)^2}{2 F^1_{G0}}}\,.
\eea
Now we note that using the explicit expressions in \rf{eq:DABC} we find
\bea\label{eq:expABC}
\left.\bar z^{\bar1}\, \Delta A\,  z^1\right|_{z^1=\frac{(\chi^1)^2}{2 F^1_{G0}}} =& \Delta C &= - \mathcal{M}\frac{ (\chi^{1})^2\, \,  (\chi^{\bar 1})^2}{4}\,,\cr
 \left.z^1  \Delta \bar B\right|_{z^1=\frac{(\chi^1)^2}{2 F^1_{G0}}}=& \left. \Delta B\bar z^{\bar 1}\right|_{z^1=\frac{(\chi^1)^2}{2 F^1_{G0}}}&=+\mathcal{M}\frac{ (\chi^{1})^2\, \,  (\chi^{\bar 1})^2}{4}\,.
\eea
This implies that
\be
\ls \bar z^{\bar1}\, \Delta A^1 \,  z^1+ z^1 \Delta \bar{B}^1 + \Delta B^1 \bar z^{\bar 1}+ \Delta C^1 \rs_{z^1=\frac{(\chi^1)^2}{2 F^1_{G0}}}= 0\,.
\ee
We also see from \rf{eq:expABC} that 
\be
\ls \bar z^{\bar1}\, \Delta A\,  z^1 + \Delta B\bar z^{\bar 1}\rs_{z^1=\frac{(\chi^1)^2}{2 F^1_{G0}}}=0\,.
\ee
Thus we have explicitly shown that the Lagrangian \eqref{eq:Lapp} reduces to the one in equation \eqref{eq:L5}.

\section{Inverse {\K} metric and Christoffel symbols}\label{app:C}
From the {\K} metric as given in eqn. \rf{eq:K} we trivially find the following expansion in $z^1$ and $\bar{z}^{\bar 1}$
\bea
g_{1 \bar 1} &=& g_{1 \bar 1}(z^i,\bar{z}^{\ib})\,,\cr
g_{1 \ib} &=&  \bar z^{\bar 1} \partial_\ib g_{1 \bar 1}\,,\cr
g_{i \bar 1} &=& z^{1} \partial_i g_{1 \bar 1}\,,\cr
g_{i \jb} &=& K_{0,i\jb} + z^1 \bar z^{\bar 1} \partial_i \partial_\jb g_{1 \bar 1}\,.
\eea
Contracting the {\K} metric $g_{\alpha \beta}$ with its inverse we find the following relations
\bea\label{eq:gginv}
g_{1 \bar 1} g^{\bar 1 1}+g_{1 \jb} g^{\jb 1} &=& 1\,,\cr
g_{1 \bar 1} g^{\bar 1k}+g_{1 \jb} g^{\jb k} &=& 0\,,\cr
g_{i \bar 1} g^{\bar 1 1}+g_{i \jb} g^{\jb 1} &=& 0\,,\cr
g_{i \bar 1} g^{\bar 1 k}+g_{i \jb} g^{\jb k} &=& {\delta_i}^k\,.
\eea
Now we recall that
\be
(g_{\jb i})^{-1}=  (K_{0,\jb i})^{-1}-z^1 \bar z^{\bar 1}  (K_{0,\jb l})^{-1}  (K_{0,i \bar m})^{-1} \partial_l \partial_{\bar m} g_{1 \bar 1}\,,
\ee
and use the above \rf{eq:gginv} to find the expansion of the inverse metric
\bea\label{eq:ginv}
g^{1 \bar 1} &=& \frac{1}{g_{1 \bar 1}-g_{1 \jb} (g_{i \jb})^{-1} g_{i \bar 1}} = \frac{1}{g_{1 \bar 1}}+ z^1 \bar{z}^{\bar 1}\frac{ (\partial_\jb g_{1 \bar 1}) (K_{0,i\jb})^{-1}  (\partial_i g_{1 \bar 1})}{(g_{1 \bar 1})^2}\,,\cr
g^{1 \jb} &=& -(g_{i \jb})^{-1} g_{i \bar 1} g^{1 \bar 1} =- z^1 \frac{(K_{0,i\jb})^{-1} (\partial_i g_{1 \bar 1})}{g_{1 \bar 1}}\,,\\
g^{k\jb} &=& (g_{i\jb})^{-1} \lp {\delta_i}^k - g_{i \bar 1 } g^{\bar 1 k} \rp\cr
&=&  (K_{0,k\jb})^{-1}-z^1 \bar z^{\bar 1} \ls (K_{0,\jb l})^{-1}  (K_{0,k \bar m})^{-1} \partial_l \partial_{\bar m} g_{1 \bar 1} - \frac{(K_{0,i\jb})^{-1}(\partial_i g_{1 \bar 1}) (K_{0,k \bar m})^{-1} (\partial_{\bar m} g_{1 \bar 1})}{g_{1 \bar 1}})  \rs\,.\nn
\eea

The Christoffel symbols for the {\K} manifold are given by 
$\Gamma^\alpha_{\beta \gamma} = g^{\alpha \bar \delta} \partial_\beta g_{\gamma \bar \delta}$. 
For the Christoffel symbols with upper index $\alpha= 1$ we explicitly find
\bea\label{eq:Christoffel}
\Gamma^{1}_{11} &=& g^{1 \bar 1} \partial_1 g_{1 \bar 1}+ g^{1 \ib} \partial_1 g_{1 \ib}= 0\,,\cr
\Gamma^{1}_{1i} &=&  g^{1 \bar 1} \partial_1 g_{i \bar 1}+ g^{1 \jb} \partial_1 g_{i \jb}\cr
&=& \frac{\partial_i g_{1 \bar 1}}{g_{1 \bar 1}} + z^1 \bar z^{\bar 1}\lp \frac{(\partial_i g_{1 \bar 1}) (\partial_\jb g_{1 \bar 1}) (K_{0,k\jb})^{-1}  (\partial_k g_{1 \bar 1})}{(g_{1 \bar 1})^2} -\frac{(\partial_i \partial_\jb g_{1 \bar 1})(K_{0,k\jb})^{-1} (\partial_k g_{1 \bar 1})}{g_{1 \bar 1}} \rp\,,\cr
\Gamma^{1}_{ij} &=&  g^{1 \bar 1} \partial_i g_{j \bar 1}+ g^{1 \bar k} \partial_i g_{j \bar k}= z^1 \lp \frac{\partial_i \partial_j g_{1 \bar 1}}{g_{1 \bar 1}}-\frac{(K_{0,l\bar k})^{-1} (\partial_l g_{1 \bar 1})}{g_{1 \bar 1}} K_{0,ij\bar k}\rp\,.
\eea

\section{The expansion of $F^1_{G}$}\label{app:F1}
Using the equations from the previous appendix we can now expand $F^1_{G}$ to get the explicit expression for $F^1_{G0}=F^1_{G}|_{z^1=\bar z^{\bar 1}=0}$. We will also show that $F^1_{G\bar 1}=0$, i.e. that $F^1_G$ has no terms that only depend on $\bar z^{\bar 1}$ and not on $z^1$. It then follows automatically that its complex conjugate $\bar F^{\bar 1}_{G 1}$ vanishes as well.

Let us first recall the definition of $F^1_G$   in eqns. \eqref{eq:F} and \eqref{eq:Ff},
\be
F^1_G = -\rme^{\frac{K}{2}} g^{1 \bar \beta} \overline\nabla _{\bar \beta }\overline{W} + \frac12 \Gamma^1_{\beta \gamma} \bar \chi^\beta \chi^\gamma +\frac14 \bar f_{AB \bar{\beta}} g^{\bar \beta 1} \bar{\lambda}^A P_R \lambda^B\,.
\ee
Expanding the bosonic part in powers of $z^1$ and $\bar z^{\bar 1}$ we find using \eqref{eq:ginv} that
\bea\label{eq:FG1}
&&\rme^{\frac{K}{2}}\lp g^{1 \bar 1} \overline\nabla _{\bar1 }\overline{W}+ g^{1 \ib} \overline\nabla _{\ib }\overline{W}\rp \cr
&=& \rme^{\frac{K_0}{2}}\lp 1 +\frac{z^1 \bar{z}^{\bar 1} g_{1 \bar1}}{2}\rp\lp g^{1 \bar 1} (\overline{W}_{\bar 1}+z^1 g_{1 \bar1} \overline{W})-z^1\frac{(K_{0,j\ib})^{-1} (\partial_j g_{1 \bar 1})}{g_{1 \bar 1}} \overline\nabla _{\ib }\overline{W}\rp\cr
&=&  \frac{\rme^{\frac{K_0}{2}}\,\overline{W}_{\bar 1}}{g_{1 \bar 1}} + \mathcal{O}(z^1,z^1 \bar{z}^{\bar 1})\,.
\eea
Since $\overline{W}_{\bar 1}$ is independent of  $\bar z^{\bar 1}$ we see that the above expression contains no term linear in $\bar z^{\bar 1}$ (and independent of $z^1$).

The term $\frac12 \Gamma^1_{\beta \gamma} \bar \chi^\beta \chi^\gamma$ can only have a term linear in $\bar z^{\bar 1}$, if $\Gamma^1_{\beta \gamma}$ has such a term. However, from the explicit expressions in equation \eqref{eq:Christoffel} we see that this is not the case. In particular, we find the expansion
\be\label{eq:FG2}
\frac12 \Gamma^1_{\beta \gamma} \bar \chi^\beta \chi^\gamma = \frac{(\partial_i g_{1 \bar 1})\bar \chi^{1} \chi^i}{2 g_{1 \bar 1}}  + \mathcal{O}(z^1,z^1 \bar{z}^{\bar 1})\,.
\ee 
Finally, we see that the gaugino term can be expanded as
\be\label{eq:FG3}
\frac14 \bar f_{AB \bar{\beta}} g^{\bar \beta 1} \bar{\lambda}^A P_R \lambda^B = \frac14 \lp \bar f_{AB \bar{1}} g^{\bar 1 1} + \bar f_{AB \ib} g^{\ib 1}\rp  \bar{\lambda}^A P_R \lambda^B  =\frac{ \bar f_{AB \bar{1}} \bar{\lambda}^A P_R \lambda^B}{4 g_{1 \bar 1}}  + \mathcal{O}(z^1,z^1 \bar{z}^{\bar 1})\,,
\ee
and does not have a linear term in $\bar z^{\bar 1}$ either. So we conclude that $F^1_G$ has no linear terms in $\bar z^{\bar 1}$ and therefore by definition we have $F^1_{G \bar 1}=0$.

Gathering the terms in \rf{eq:FG1}-\rf{eq:FG3} we find
\be
F^1_{G0} =  \frac{1}{g_{1 \bar 1}} \lp -\rme^{\frac{K_0}{2}}\,\overline{W}_{\bar 1}+ \frac{1}{2} (\partial_i g_{1 \bar 1})\bar \chi^{1} \chi^i +\frac{1}{4 } \bar f_{AB \bar{1}} \bar{\lambda}^A P_R \lambda^B \rp\,.
\ee
As discussed above, in the action $F^1_{G0}$ always appears multiplied by $(\chi^1)^2 = \bar \chi^1 \chi^1$ like for example in 
\be
z^1_G=\frac{(\chi^1)^2}{2 F^1_{G0}}=-\frac{g_{1 \bar 1}(\chi^1)^2}{2  \rme^{\frac{K_0}{2}}\,\overline{W}_{\bar 1}} \sum_{n \geq 0} \lp\frac{\frac{1}{2} (\partial_i g_{1 \bar 1})\bar \chi^{1} \chi^i +\frac{1}{4 } \bar f_{AB \bar{1}} \bar{\lambda}^A P_R \lambda^B}{ \rme^{\frac{K_0}{2}}\,\overline{W}_{\bar 1}}\rp^n\,.
\ee
Since $(\chi^1)^2 \bar \chi^{1} \chi^i=0$ we can drop the term linear in $\chi^1$ and find 
\be
z^1_G=\frac{(\chi^1)^2}{2 F^1_{G0}}=-\frac{g_{1 \bar 1}(\chi^1)^2}{2  \rme^{\frac{K_0}{2}}\,\overline{W}_{\bar 1}}\sum_{n \geq 0} \lp\frac{\frac{1}{4 } \bar f_{AB \bar{1}} \bar{\lambda}^A P_R \lambda^B}{ \rme^{\frac{K_0}{2}}\,\overline{W}_{\bar 1}}\rp^n \equiv \frac{(\chi^1)^2}{2 f^1}\,,
\ee
where we defined
$
f^1\equiv    \frac{1}{g_{1 \bar 1}} \lp- \rme^{\frac{K_0}{2}}\,\overline{W}_{\bar 1}+\frac{1}{4 } \bar f_{AB \bar{1}} \bar{\lambda}^A P_R \lambda^B \rp
$
to contain all the terms in $F^1_{G0}$ that are independent of $\chi^1$.


\end{document}